Feasibility-Preserving Quantum Search for Constrained Transportation Routing

Dahye Kim[a], Monika Filipovska[a, *]

[a] School of Civil and Environmental Engineering, University of Connecticut
* Corresponding author, monika.filipovska@uconn.edu

---

## Abstract
Transportation routing problems such as the Traveling Salesperson Problem (TSP) and the Vehicle Routing Problem (VRP) are characterized by strict feasibility requirements involving customer assignment and visit rules, route sequencing, and depot-return logic alongside cost minimization. Most quantum routing formulations adopt Quadratic Unconstrained Binary Optimization (QUBO) encodings, where feasibility is incorporated indirectly via penalty terms in the cost Hamiltonian. While convenient for standard implementations of the Quantum Approximate Optimization Algorithm (QAOA), QUBO encodings allow the quantum search dynamics to allocate substantial probability to infeasible route configurations. This study develops a transportation-grounded constraint-aware Quantum Alternating Operator Ansatz ($QAOA^+$) framework that embeds feasibility-preserving logic directly into the search operator. We introduce a custom mixer that functions as a quantum analogue of feasibility-preserving routing neighborhoods, using column-wise swap moves, it restricts evolution to feasible configurations while enabling structured exploration of valid routes. We compare three constraint-handling architectures: penalty-based QUBO-QAOA, penalty-free $QAOA^+$ with the feasibility-preserving mixer, and a Hybrid $QAOA^+$ combining mixer-based feasibility with and penalty guidance. Results on small TSP and VRP instances show that constraint-handling architecture strongly influences feasible-route sampling, convergence behavior, and probability concentration over low-cost feasible routes. These findings position constraint-aware quantum search as a methodological extension of transportation routing search approaches, where feasibility is enforced through admissible quantum transitions rather than post-hoc penalties.

---

## 1. Introduction and Motivation

Modern transportation and logistics systems increasingly rely on optimization to support routing, scheduling, and other operational decisions under tight service and resource constraints (Borndörfer et al., 1998). In applications such as vehicle routing, logistics planning, and fleet operations, a candidate route is useful only if it satisfies required feasibility constraints. These feasibility requirements make routing problems fundamentally different from unconstrained combinatorial optimization problems, and constraint handling is a decisive factor for both optimization efficiency and solution quality (Adamo et al., 2024; Zhou et al., 2023). Therefore, how feasibility is represented and enforced is a central methodological issue in transportation routing optimization. The Traveling Salesperson Problem (TSP) and the Vehicle Routing Problem (VRP) provide canonical testbeds for examining this issue because they combine cost minimization with strict sequencing, assignment, and depot-return requirements (Dantzig and Ramser, 1959). These problems have long been recognized as NP-hard (Karp, 1972; Lenstra and Kan, 1981), and transportation applications have therefore relied extensively on heuristics and metaheuristics to obtain high-quality feasible solutions within reasonable computational time (Deb et al., 2016; Sidorov and Morozov, 2021).

Classical transportation optimization has developed a rich set of mechanisms for handling routing feasibility. Exact formulations impose feasibility through linear constraints, subtour-elimination structures, and vehicle-flow consistency conditions. Heuristic and metaheuristic methods, in contrast, often rely on problem-specific route construction rules, neighborhood moves, and repair operators to avoid or correct infeasible solutions (Tan and Yeh, 2021). In this sense, feasibility is not treated as an afterthought in classical routing methodology; it is built into the way the algorithm represents and constructs routes. As quantum optimization is introduced into transportation routing, a key methodological question is whether such feasibility logic can be embedded into the architecture of quantum search itself, rather than enforced only through external penalties or post-processing.

Most existing quantum applications to TSP and VRP adopt a Quadratic Unconstrained Binary Optimization (QUBO) formulation, where routing constraints are converted into penalty terms in the cost function. This strategy is convenient because it makes the problem compatible with standard Quantum Approximate Optimization Algorithm (QAOA) and quantum annealing implementations. Prior studies have demonstrated proof-of-concept applications of these algorithms to small-scale routing problems (Azad et al., 2023; Cattelan and Yarkoni, 2024). However, penalty-based QUBO formulations do not prevent the algorithm from generating route configurations that violate basic transportation feasibility requirements. As a result, part of the sampled probability distribution may be assigned to routes with duplicated customers, missing customers, invalid vehicle assignments, or discontinuous route structures. From a transportation modeling perspective, this is not only a computational inefficiency but also a methodological limitation, because the search process may allocate substantial sampling effort outside the feasible routing space that defines the problem.

The Quantum Alternating Operator Ansatz ($QAOA^+$) provides an alternative framework for constrained optimization by allowing feasibility requirements to be incorporated directly into the search process through a mixer operator (Hadfield et al., 2019). In QAOA, the mixer determines how the quantum state moves among encoded solution configurations. A generic mixer may move through both feasible and infeasible states, whereas a problem-specific mixer can restrict transitions to valid states. Despite this potential, tailored mixer constructions for routing problems such as TSP and VRP remain limited, and existing studies have focused primarily on general quantum optimization frameworks or penalty-based encodings rather than systematic comparisons of how different constraint-handling architectures affect transportation routing search behavior (Bourreau et al., 2025; Fuchs et al., 2022).

This study addresses this gap by developing a transportation-grounded, constraint-aware quantum search framework for TSP and VRP. Specifically, we design a custom mixer that embeds one-visit, one-customer-per-position, and vehicle-assignment logic directly into the quantum search architecture. The proposed mixer follows a column-wise structure over vehicle-route positions, allowing the quantum evolution to remain within the feasible routing subspace. This enables us to examine how operator-level feasibility enforcement changes the sampled routing distribution compared with a conventional penalty-based QUBO-QAOA formulation.

To isolate the methodological effect of constraint handling, we compare three formulations: a conventional QUBO-based QAOA that relies on penalty terms, a penalty-free $QAOA^+$ formulation that employs the proposed feasibility-preserving mixer, and a Hybrid $QAOA^+$ formulation that combines mixer-level constraint enforcement with penalty-based guidance. Because current quantum hardware and circuit simulation remain limited in scale, the computational experiments are designed as controlled small-instance studies. The purpose is not to claim large-scale computational superiority, but to evaluate how constraint-handling architecture affects feasibility preservation, sampling efficiency, convergence behavior, and the distribution of high-quality feasible routing solutions.

The contributions of this study are threefold. First, it develops a transportation-grounded $QAOA^+$ framework in which routing feasibility requirements are translated into feasibility-preserving mixer operations rather than imposed only as cost penalties. Second, it formalizes a transportation-grounded constraint-handling architecture by showing how feasibility-preserving routing neighborhood logic can be implemented as admissible quantum transitions, and by clarifying the architectural trade-off between mixer-level feasibility preservation and penalty-based feasibility treatment in quantum optimization. Third, it provides a controlled computational comparison of penalty-based QUBO-QAOA, penalty-free $QAOA^+$, and Hybrid $QAOA^+$ formulations on TSP and VRP instances, showing how different constraint-handling architectures affect feasibility preservation, convergence behavior, and the probability distribution over high-quality feasible routes. Together, these contributions clarify how quantum optimization can augment existing transportation routing approaches when its operators are designed to preserve transportation-specific constraint structure.

The remainder of this paper is organized as follows. Section 2 reviews the literature on transportation routing methods, quantum optimization, and constraint handling, and identifies the research gap addressed in this study. Section 3 presents the proposed constraint-aware QAOA$^{+}$ framework, including the classical routing formulation, quantum encoding, cost Hamiltonian, and feasibility-preserving mixer design. Section 4 reports the computational experiments and comparative results for TSP and VRP instances, followed by a discussion of methodological implications for transportation routing optimization. Section 5 concludes the paper and outlines directions for future research.

## 2. Background and Related Methodological Literature

### *2.1. Structural Constraint Handling in Transportation Routing*

Transportation routing problems are not defined solely by cost minimization, but also by the need to construct routes that satisfy operational feasibility requirements. Recent transportation routing studies show that these feasibility requirements have become increasingly structural as routing models incorporate service consistency, path consistency, synchronization, grouping restrictions, vehicle-specific operations, and stochastic demand conditions, and this structural dependence directly shapes how routing methods enforce constraints. For example, the consistent vehicle routing problem on road networks explicitly incorporates path-consistency requirements through a coupled two-layer network and a branch-price-and-cut framework (Yao et al., 2021). Similarly, two-echelon vehicle routing with grouping constraints and simultaneous pickup and delivery uses a tight path-based model, customized valid inequalities, and an exact branch-cut-and-price approach to enforce service consistency across satellites and customer groups (Li et al., 2022). More recently, stochastic consistent vehicle routing has been formulated as a two-stage scenario-based model in which customer-driver assignments remain fixed while customer realizations and routes vary across scenarios (Alvarez et al., 2024). These examples show that routing feasibility is increasingly encoded as structure, and thus the methodological question becomes how search algorithms represent and preserve that structure.

Methodologically, the literature has developed three broad ways of handling routing feasibility: exact methods based on mathematical programming, metaheuristics and neighborhood-based search, and strategies that regulate controlled violations through penalties and repair. In exact methods, feasibility is typically embedded into the mathematical formulation and strengthened through algorithmic components such as decomposition, branching rules, and separation routines. Branch-and-cut methods enforce route feasibility by adding problem-specific constraints and cuts that eliminate infeasible route structures, while branch-price-and-cut methods generate feasible route columns through pricing subproblems that incorporate resource, sequencing, and operational restrictions. For example, the traveling salesman problem with multiple drones has been solved using a branch-and-cut algorithm with a new mixed-integer linear programming (MILP) formulation, and subtour-elimination constraints (Cavani et al., 2021). A two-echelon vehicle routing problem with drones has also been addressed through both MILP, with a branch-and-price algorithm used to handle allocation and routing decisions across vehicles and drones (Zhou et al., 2023). In vehicle routing with load-dependent drones, nonlinear energy-consumption feasibility is incorporated into a mixed-integer programming model and solved through a branch-price-and-cut algorithm (Xia et al., 2023). In other words, exact routing methods implement feasibility by constraining the search space itself through cuts, branching, and feasibility-aware pricing. The methodological takeaway is that feasibility enforcement in routing is not just a modeling choice, but a computational design principle realized through separation, decomposition, and route-generation rules.

Heuristic and metaheuristic routing methods follow a related logic by controlling how the search moves through the solution space. Instead of freely exploring arbitrary binary configurations, these methods use route-construction rules, neighborhood moves, and local search procedures that either preserve feasibility or restore it after controlled perturbations. For instance, the vehicle routing problem with delivery options has been solved using a large neighborhood search heuristic coupled with a set-partitioning model, where the operators are selected and evaluated according to their ability to construct feasible combinations of delivery locations and vehicle routes (Dumez et al., 2021). Last-mile delivery under stochastic customer availability and multiple visits has been addressed using a parallel adaptive large neighborhood search framework, where customer visit attempts are removed and reinserted into feasible routes while penalties are used as recourse actions for failed deliveries (Özarık et al., 2023). A soft-clustered capacitated arc routing problem has also been solved through a bilevel hybrid iterated search method, where the algorithm is designed around the constraint that all required edges in the same cluster must be served by the same vehicle (Zhou et al., 2024). Methodologically, the key point is that neighborhood design functions as an implicit constraint system by determining which route transformations are admissible and therefore which parts of the feasible region are reachable during search.

A complementary class of transportation routing methods allows limited violations or uncertain feasibility outcomes but manages them through soft constraints and penalty terms. This is particularly visible in routing problems with soft service requirements, stochastic travel times, and uncertain customer availability. For example, heterogeneous fleet routing with soft time deadlines allows tardiness at penalty cost and develops compact formulations, valid inequalities, and branch-and-cut schemes to manage the interaction between vehicle-specific routing and deadline-related feasibility (Han and Yaman, 2024). The multi-visit drone-assisted routing problem with soft time windows and stochastic truck travel times is formulated as a two-stage stochastic model and solved using a hybrid metaheuristic with sample average approximation and rolling-horizon decision logic to reduce time-window violations (Meng et al., 2024). Driver routing and scheduling with synchronization constraints has similarly been formulated on a time-expanded directed multigraph and solved with a metaheuristic enriched by destructive bound improvement, reflecting the need to manage synchronization feasibility in long-distance bus operations (Ammann et al., 2023). These methods treat feasibility as a tunable design choice, where algorithm performance depends on how violations are quantified and corrected during search rather than simply filtered out afterward.

Taken together, across exact methods, neighborhood-based heuristics, and penalty/recourse strategies, the routing literature consistently treats feasibility as a property that must be engineered into the search architecture. When routing is mapped to quantum optimization, the analogous methodological challenge is to ensure that quantum evolution respects this same feasibility structure, not merely that the objective function can be expressed in a Hamiltonian. This motivates the constraint-aware QAOA$^{+}$ framework developed in this study, which embeds routing feasibility directly into the design of the mixing operator so that search transitions remain within valid routing solutions rather than bring steered indirectly through penalty terms in a QUBO objective.

*2.2. Quantum Algorithms as Emerging Search Architectures for Transportation Optimization*

The increasing structural complexity of constrained routing problems has motivated interest in alternative computational architectures for transportation optimization. Quantum computing has recently attracted attention in this context because it offers a probabilistic search architecture for manipulating amplitudes over encoded solution configurations, providing a different computational representation of combinatorial search spaces (Feynman, 1982; Schumacher, 1995). At a basic level, quantum algorithms use qubits, which can represent superpositions of classical binary states, and quantum operations, which transform probability amplitudes over encoded solution configurations. For transportation optimization, the relevance of this representation is not that quantum computing automatically guarantees superior performance, but that it provides a different way to represent and manipulate large combinatorial search spaces (Nielsen and Chuang, 2010; Symons et al., 2023).

Early transportation-oriented quantum studies have begun to examine this potential across traffic optimization, network design, logistics, and routing. Recent surveys describe quantum computing as an emerging computational paradigm for intelligent transportation systems, while transport network design has been formulated as a Quadratic Unconstrained Binary Optimization (QUBO) problem and solved using quantum annealing as a proof-of-concept application in transportation planning (Dixit and Niu, 2023; Zhuang et al., 2024). Quantum optimization has also been applied to vehicle routing variants, including heterogeneous vehicle routing, where QAOA-based formulations map routing decisions to Ising Hamiltonians and evaluate small-size problem instances under current quantum hardware and simulation limitations (Fitzek et al., 2024). These studies show that transportation problems can be reformulated within quantum-compatible search architectures, motivating closer attention to how such architectures represent routing feasibility.

However, for routing applications, quantum optimization must address feasibility as a first-order methodological requirement, not an afterthought. As discussed in *Section 2.1*, classical routing methods enforce feasibility by controlling admissible states and transitions during search, using constraints and cuts, neighborhood operators, and penalty/repair mechanisms. Current Noisy Intermediate-Scale Quantum (NISQ) further amplify the importance of this issue due to limits on qubit counts, circuit depth, and noise sensitivity, making inefficient exploration of infeasible solutions especially costly. Therefore, the key methodological challenge is not merely mapping transportation routing problems onto quantum hardware, but embedding search feasibility logic into quantum architectures similar to the classical methods we are familiar with.

### *2.3. Constraint Handling in Quantum Optimization for Routing*

Quantum optimization methods typically require constrained transportation routing problems to be encoded into a form that can be represented and evolved by a quantum circuit or quantum annealer. The most common approach is to reformulate the constrained problem as a Quadratic Unconstrained Binary Optimization (QUBO) model, where feasibility requirements are incorporated into the objective function through penalty terms. This approach is attractive because it allows routing problems to be mapped onto standard quantum optimization frameworks, including quantum annealing and the Quantum Approximate Optimization Algorithm (QAOA). In QAOA, the algorithm alternates between a cost operator that encodes the objective function and a mixer operator that drives exploration across encoded solution states (Farhi et al., 2014). However, in the standard formulation, the mixer is generally not aware of the routing constraints. Feasibility is therefore indirectly enforced through penalties, rather than structurally preserved by the search dynamics itself.

Existing quantum routing studies largely follow this penalty-based strategy. Routing and ride-hailing problems have been formulated as QUBO models to make them compatible with quantum annealing or QAOA-type solvers, and recent applications to pickup-and-delivery and ride-hailing problems demonstrate how transportation constraints can be translated into binary variables and penalty terms (Azad et al., 2023; Bourreau et al., 2024; Cattelan and Yarkoni, 2024; Curuliuc and Leon, 2026). This line of work is valuable because it shows that transportation routing problems can be encoded for quantum optimization. However, from a transportation network modeling perspective, penalty-based QUBO encodings impose a fundamentally different constraint-handling logic than the structural mechanisms used in classical routing algorithm, such as cuts, route-generation, and feasibility-preserving neighborhoods (Xia et al., 2023; Yao et al., 2019). As a result, QUBO formulations, while quantum-compatible, still permit substantial exploration of infeasible routing solutions during the search and would benefit from the feasibility-preserving search architectures that underlie much of the classical routing literature.

The Quantum Alternating Operator Ansatz ($QAOA^{+}$) provides a mechanism to address this limitation by allowing constraints to be incorporated into the search operator itself. Rather than relying only on penalty terms in the cost function, $QAOA^{+}$ generalizes QAOA by allowing problem-specific mixing operators that can restrict evolution to a desired feasible subspace (Hadfield et al., 2019). Constraint-preserving mixer frameworks further show that quantum evolution can be restricted to subspaces satisfying hard constraints, including one-hot type constraints that are directly relevant to routing encodings (Fuchs et al., 2022). This capability is particularly important for transportation routing, where infeasible states may correspond to duplicated customer visits, missing customers, invalid vehicle assignments, or discontinuous routes. However, routing-specific $QAOA^{+}$ mixer designs remain far less developed than QUBO-based encodings, and transportation-grounded evidence is still limited on how penalty-based versus mixer-based architectures affect feasible search behavior and solution sampling in routing problems.

### *2.4. Research Gaps*

The preceding literature identifies a gap between the way feasibility is treated in transportation routing methodology and the way it is commonly represented in quantum routing models. Classical routing methods typically embed feasibility into formulations, state representations, route-generation mechanisms, neighborhoods, or repair operators, whereas most quantum routing studies rely on QUBO-based penalty formulations that penalize infeasible solutions without structurally restricting the search dynamics. Although $QAOA^{+}$ provides a framework for embedding constraints directly into the mixer operator, routing-specific feasibility-preserving mixer designs remain limited, and systematic transportation-grounded comparisons between penalty-based and mixer-based constraint-handling architectures are still lacking.

Therefore, the gap addressed in this study is not simply the lack of quantum applications to TSP and VRP. Rather, it is the lack of methodological guidance on how feasibility-preserving search principles from transportation routing can be translated into quantum search architecture, and how this architectural choice affects feasibility preservation, convergence behavior, and the sampled distribution of high-quality routes.

### *2.5. Contributions*

This study addresses the above gap by developing and evaluating a constraint-aware QAOA$^+$ architecture for TSP and VRP that is grounded in feasibility-preserving search principles from transportation routing. The main contributions are as follows:

(1) We translate routing feasibility requirements into a feasibility-preserving quantum mixer design. Specifically, one-visit, one-customer-per-position, and vehicle-assignment logic are embedded into the quantum search operator, enabling the quantum search process to remain confined to feasible solution subspaces.

(2) We formalize a transportation-grounded constraint-handling architecture by showing how feasibility-preserving routing neighborhood logic can be implemented as admissible quantum transitions and clarify the architectural trade-off between mixer-level and penalty-based feasibility treatment in quantum optimization.

(3) We conduct a controlled computational comparison of three constraint-handling architectures, namely penalty-based QUBO-QAOA, penalty-free QAOA$^+$, and Hybrid QAOA$^+$, to examine how penalty-based and mixer-based mechanisms affect feasibility preservation, convergence behavior, and the probability distribution over high-quality feasible routes.

Together, these contributions position quantum optimization not as a generic application of a quantum algorithm to routing, but as a methodological extension of transportation routing logic in which feasibility-preserving search is implemented through quantum operator design.

## 3. Methodological Framework

Building on the research gap identified in *Section 2*, this section develops the proposed constraint-aware QAOA$^+$ framework by translating transportation routing feasibility requirements into quantum encoding, cost Hamiltonian design, and mixer operations. In this study, QAOA and QAOA$^+$ are interpreted as alternative search architectures over encoded route configurations. The key methodological distinction is how each architecture handles routing feasibility. Standard QAOA relies primarily on penalty terms in the cost Hamiltonian to discourage infeasible route configurations, whereas QAOA$^+$ embeds feasibility into the mixer, which defines allowable transitions among route states. This distinction directly reflects the transportation routing issue discussed earlier: whether feasibility is treated only as a cost penalty or built into the structure of the search process. For more technical details, readers are encouraged to reference the original papers for QAOA (Farhi et al., 2014) and QAOA$^+$ (Hadfield et al., 2019).

### *3.1. QAOA and QAOA$^+$ as Routing Search Architectures*

QAOA is a hybrid quantum-classical algorithm for approximate combinatorial optimization (Farhi et al., 2014). It alternates between two operators: a cost Hamiltonian, which assigns phase information according to the objective value of a candidate solution, and a mixer Hamiltonian, which moves the quantum state across encoded solution configurations. QAOA solves optimization by formulating problems as QUBO (Quadratic Unconstrained Binary Optimization) where constraints are incorporated as penalty terms in the cost Hamiltonian (Glover et al., 2018). Since constraints are included in the cost Hamiltonian, the mixer Hamiltonian typically plays a generic role (e.g., a simple transverse field mixer $\sum X_i$ ) of flipping qubits for exploration, without explicit awareness or enforcement of the problem constraints. For a routing problem, the cost Hamiltonian is therefore constructed to represent the total travel cost, while the mixer Hamiltonian serves to explore encoded route configurations.

The algorithm starts by placing the qubits into a superposition of all possible solutions, then evolves the system by repeatedly applying the cost Hamiltonian to bias the state towards optimal solutions and the mixer Hamiltonian to explore new configurations. These cost and mixer operations are repeated over multiple layers. Increasing the number of layers can improve the expressive power of the ansatz, but it also increases circuit depth and noise exposure on near-term quantum devices (Harrigan et al., 2021). After the cost and mixer operators are applied, the circuit is measured to produce a classical bitstring representing a candidate route configuration. Because quantum measurement is probabilistic, the circuit is executed many times, or shots, to estimate a sampling distribution over route configurations. This distribution is then used to compute expected objective values and to guide the classical optimizer in updating the QAOA parameters. The overall process of QAOA is described in Figure 1.

However, this generic exploration creates a limitation for constrained transportation routing problems. If feasibility is represented only through penalties, infeasible configurations may still receive sampling probability during the search. QAOA$^+$ provides a mechanism to address this issue by replacing the generic mixer with a problem-specific, constraint-aware mixer (Hadfield et al., 2019). In the context of this study, the QAOA$^+$ mixer is designed as a feasibility-preserving routing neighborhood operator: it allows transitions among encoded route states only when those transitions preserve the core routing requirements, such as one-customer-per-position and one-visit-per-customer logic. The conceptual difference between standard QAOA and QAOA+ is shown in Figure 2.

This distinction is summarized in Table 1. It clarifies how different constraint-handling architectures translate into different routing search behaviors. In QUBO-based QAOA, feasibility is encouraged indirectly through penalty terms in the cost Hamiltonian. In contrast, in QAOA$^+$, feasibility is built into the mixer so that the search dynamics are structured around valid route transitions. This transportation interpretation motivates the formulation developed in the following subsections.

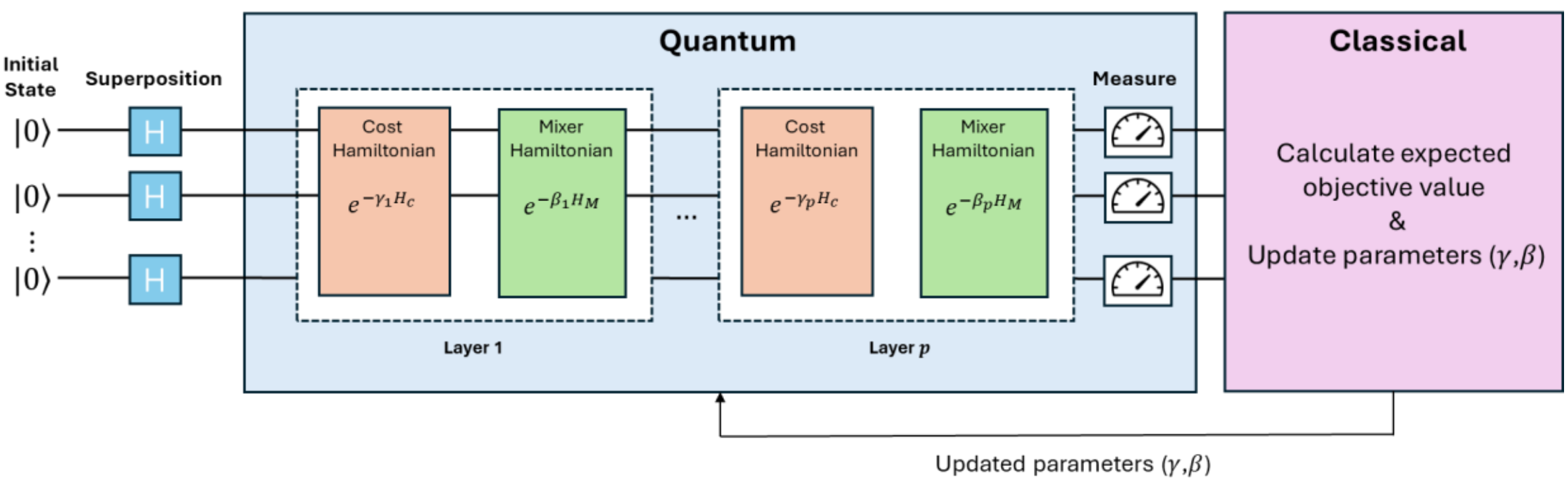

**Figure 1 Schematic structure of the Quantum Approximate Optimization Algorithm (QAOA), illustrating the alternating application of the cost Hamiltonian and mixer Hamiltonian across multiple circuit layers, with parameters optimized through a classical outer loop**

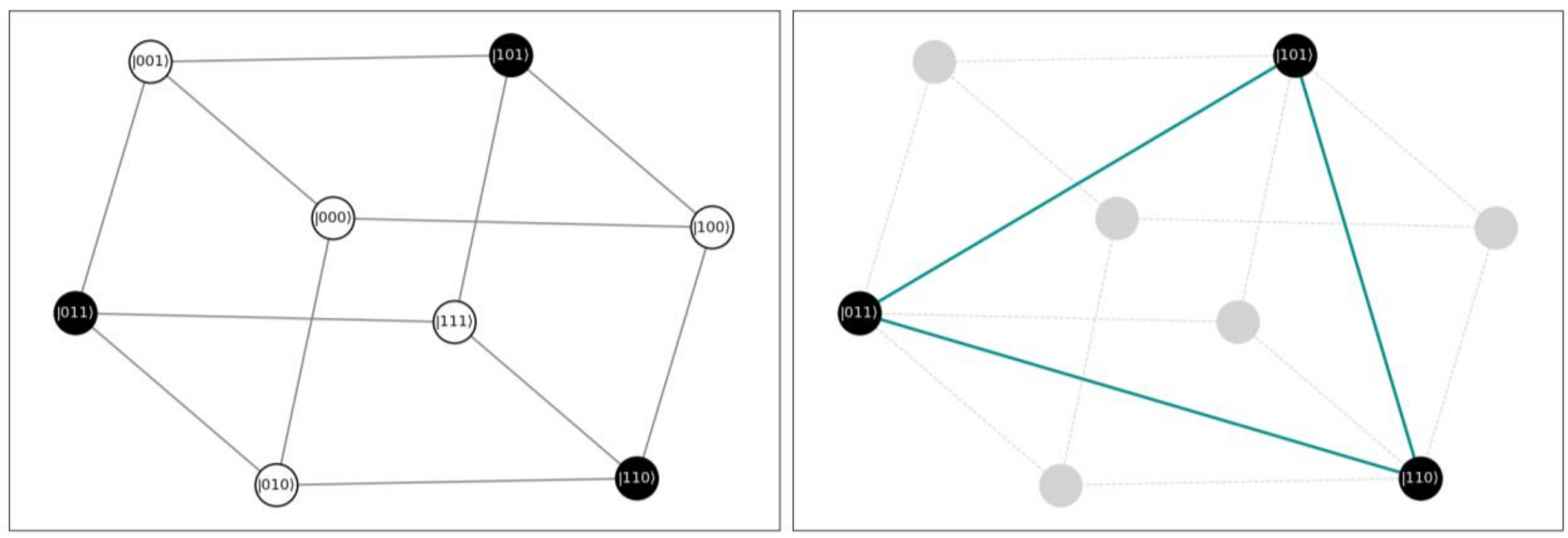

**Figure 2 Conceptual comparison of Mixer Operations in standard QAOA and QAOA+. In standard QAOA (left), the generic mixer explores the entire solution space, connecting both feasible (black) and infeasible (white) states. In QAOA$^+$ (right), a constraint-preserving mixer restricts transitions to feasible subspace.**

**Table 1 Comparative Summary of QAOA and QAOA⁺ Features**

| Feature | QAOA | QAOA⁺ |
|---|---|---|
| **Constraint** | Via penalty terms added to the cost Hamiltonian | Directly through a constraint-aware mixer Hamiltonian |
| **Search Space** | May explore both feasible and infeasible route configurations | Restricted to feasible route transitions by design |
| **Mixer Design** | Generic simple mixer | Feasibility-preserving routing neighborhood operator |
| **Initial State** | Uniform superposition of encoded states | Feasible route state or superposition of feasible states |

Although the proposed QAOA⁺ mixer is motivated by feasibility-preserving routing neighborhoods in classical transportation optimization, its search mechanism is methodologically distinct from classical neighborhood search. In classical heuristic and metaheuristic routing methods, a neighborhood operator typically modifies one incumbent route configuration at a time, for example by swapping, relocating, or reinserting customers while maintaining or restoring feasibility. The search therefore proceeds through a sequence of individual route updates, guided by objective evaluation, acceptance rules, or repair mechanisms. In contrast, the QAOA⁺ mixer defines allowable transitions over a Hilbert subspace spanned by feasible routing configurations. Starting from a feasible route state or a superposition of feasible route states, the mixer restricts amplitude evolution to assignment-feasible configurations, while the cost Hamiltonian redistributes phase information according to route cost. Through repeated alternation of the cost and mixer operators, probability amplitude is redistributed over multiple feasible routes before measurement.

Thus, the quantum search does not simply replicate a classical routing neighborhood. Rather, it translates feasibility-preserving neighborhood logic into a unitary amplitude-evolution mechanism that combines structured subspace restriction, probabilistic exploration, and parallel representation of feasible route configurations. This distinction clarifies how the proposed quantum approach augments classical transportation routing methodology: it carries the classical principle of avoiding invalid route configurations into the quantum setting but does so by shaping a probability distribution over feasible route states rather than by selecting a single next route in a sequential search trajectory. Therefore, the methodological contribution is not a claim of immediate large-scale computational advantage, but an operator-design framework showing how transportation-specific feasibility logic can be embedded directly into quantum search dynamics.

*3.2. Classical Routing Problem Formulation*

Before introducing the quantum formulations of the TSP and VRP, their standard mathematical forms are presented first. Both are NP-hard problems, conventionally modeled as mixed-integer programs (MIPs). The purpose of presenting the MIP is not to propose a new classical formulation, but to establish a transparent baseline based on which the quantum encodings are developed. More specifically, the classical formulation serves as a reference map that identifies which transportation routing feasibility requirements must be preserved when the problem is translated into quantum encoding, cost Hamiltonian design, and mixer operations.

Let the set of all nodes be $N = \{0, 1, \dots, n\}$, where node $0$ represents the depot and nodes $1, \dots, n$ represent customers. Let the set of vehicles be $A = \{0, 1, \dots, a-1\}$. The cost of travel from node $i$ to node $j$ is denoted by $c_{ij}$. The primary decision variable is a binary variable $x_{ij}$, where $x_{ij} = 1$ if a vehicle traverses the arc from node $i$ to node $j$, and 0 otherwise. For a subset of customers $S \subset N\backslash\{0\}$, let $r(S)$ denote the minimum number of routes required to serve the subset of customers $S$. In the uncapacitated case considered in this study, $r(S) = 1$. Subsets $S$ are used in subtour elimination constraints to prevent disconnected cycles.

The TSP seeks the shortest possible tour for a single vehicle that visits each customer exactly once and returns to the depot. The VRP generalizes this concept to a fleet of multiple vehicles, with the objective of minimizing the total combined travel cost. In the VRP, each vehicle must depart from and return to the depot, and the total number of active routes equals the fleet size. The formulation follows the classical DFJ model (Dantzig et al., 1954).

For clarity, this study focuses on the uncapacitated VRP without time windows. This scope is intentionally chosen to isolate the effect of constraint-handling architecture, rather than to claim coverage of all operational VRP variants. Capacity, time-window, and other service constraints can be incorporated through additional flow, resource, or time-indexed variables in classical formulations. In the quantum setting, they would require additional penalty terms or more elaborate feasibility-preserving mixer designs. By focusing on the core sequencing, assignment, depot-return, and route-continuity requirements, the formulation allows a controlled comparison between penalty-based and mixer-based constraint handling. The MIP formulation of TSP is as follows:

$$\min z = \sum_{i \in N} \sum_{j \in N, i \neq j} c_{ij} x_{ij} \tag{1}$$

$$s.t.$$

$$\sum_{i \in N, i \neq j} x_{ij} = 1, \quad \forall j \in N \tag{2}$$

$$\sum_{j \in N, j \neq i} x_{ij} = 1, \quad \forall i \in N \tag{3}$$

$$\sum_{i \in S} \sum_{j \in S, i \neq j} x_{ij} \leq |S| - 1, \quad \forall S \subseteq N \setminus \{0\}, \quad 2 \leq |S| \leq n - 1 \tag{4}$$

$$x_{ij} = \{0, 1\} \tag{5}$$

The objective in equation (1) minimizes the total travel cost. Constraints (2) and (3) ensure that each customer node has exactly one incoming and one outgoing arc, while constraint (4) eliminates subtours by preventing the formation of disconnected cycles among subsets of customers.

The VRP extends the DFJ formulation of the TSP by introducing depot and fleet-size constraints. The MIP formulation of the VRP is as follows:

$$\min z = \sum_{i \in N} \sum_{j \in N, i \neq j} c_{ij} x_{ij} \tag{6}$$

$$s.t.$$

$$\sum_{i \in N, i \neq j} x_{ij} = 1, \quad \forall j \in N \setminus \{0\} \tag{7}$$

$$\sum_{j \in N, j \neq i} x_{ij} = 1, \quad \forall i \in N \setminus \{0\} \tag{8}$$

$$\sum_{j \in N \setminus \{0\}} x_{0j} = |A| \tag{9}$$

$$\sum_{i \in N \setminus \{0\}} x_{i0} = |A| \tag{10}$$

$$\sum_{i \in S} \sum_{j \in N \setminus S} x_{ij} \geq r(S), \quad \forall S \subset N \setminus \{0\}, S \neq \emptyset \tag{11}$$

$$x_{ij} = \{0, 1\} \tag{12}$$

These classical constraints define the routing feasibility structure that must be represented in the quantum formulation. In the QUBO-based formulation, these requirements are translated into penalty terms in the cost Hamiltonian. In the proposed $QAOA^+$ formulation, key elements of this feasibility logic are used to define allowable transitions among encoded route states through the mixer design. This distinction is central to the methodological comparison in this study.

To clarify the relationship between classical and quantum formulations, Table 2 summarizes the differences in constraint representation across MIP, QUBO, and QAOA$^+$. In the classical MIP model, constraints are expressed as linear equations. In contrast, the QUBO formulation incorporates constraints directly into the objective function via penalty terms, trading strict feasibility for algebraic simplicity (Lucas, 2014). However, the QAOA$^+$ framework embeds key assignment-related feasibility requirements within the mixer Hamiltonian, restricting allowable transitions among encoded route states, while travel cost, depot return, and route-continuity terms are handled through the cost Hamiltonian and contiguity treatment.

**Table 2 Mapping of Routing Feasibility Requirements Across MIP, QUBO and QAOA$^+$ Representations**

| Routing Requirement | MIP | QUBO | QAOA$^+$ |
|---|---|---|---|
| **Cost Minimization** | Objective function | Cost Hamiltonian | Cost Hamiltonian |
| **Customer Visit Uniqueness** | Degree or assignment constraints | Penalty terms | Feasible mixer transitions |
| **Vehicle-Position Assignment** | Flow or assignment constraints | Penalty terms | One-hot encoding and feasible mixer transitions |
| **Route sequencing** | Flow consistency | Penalty terms | Vehicle-position encoding and feasible route transitions |
| **Route Continuity & Depot Return** | Depot and subtour constraints | Penalty terms | Cost Hamiltonian and contiguity treatment |

*3.3. Quantum Encoding of the Routing Problems*

To implement the quantum optimization problem, we use a three-dimensional vehicle–position–customer assignment encoding. For each vehicle, a customer- position assignment table is defined, and the full route configuration is obtained by stacking these tables over the vehicle dimension. Each binary variable $x_{a,k,i}$ indicates whether vehicle $a$ visits customer $i$ at route position $k$. This encoding makes the transportation interpretation explicit: the vehicle dimension determines route assignment, the position dimension determines visit order, and the customer dimension identifies the served customer.

Let $N = \{0,1,\dots,n\}$ denote the set of all nodes, where node $0$ represents the depot, and let $C = N\backslash\{0\}$ denote the set of customer nodes (nodes to be visited). For each vehicle-position pair $(a,k)$, we assign a block of $|C|$ qubits, each corresponding to one customer node. The block follows an at-most-one-hot structure: at most one customer qubit can take value 1, while the all-zero block represents an unused route position. This directly encodes the transportation requirement that a vehicle can serve at most one customer at a given route position. The depot is not explicitly represented as a qubit because each vehicle route is assumed to start from and return to the depot. Departure and return costs are instead incorporated through the cost Hamiltonian.

Given $|A|$ vehicles, and $|C|$ customers, the maximum number of customer visits for a single vehicle, $K,$ is unknown beforehand. We therefore set $K = |C|$, ensuring that every feasible route can be accommodated, as no vehicle can visit more customers than the total number of customers. We index positions as $k \in \{1,2,\dots,K\}$. For shorter routes, some allocated positions remain unused and correspond to empty slots. Since the presence of such empty slots between valid visits can distort travel cost calculation, a penalty term in the cost Hamiltonian is later introduced to consolidate all empty slots toward the end of the route. This mechanism ensures consistency across vehicle routes regardless of route length.

To clearly illustrate the encoding, consider a toy instance with two vehicles and three customers, with node 0 as the depot and $K = 3$. This instance requires $2 \times 3 \times 3 = 18$ qubits. The bitstring is constructed by concatenating the customer-position table for Vehicle 1 followed by the table for Vehicle 2, following a lexicographic ordering (Neukart et al., 2017). For example, a feasible solution where Vehicle 1 visits Customer 1 and then Customer 2, while Vehicle 2 visits Customer 3, is encoded as 100010000 001000000. The first segment, 100010000, represents Vehicle 1 visiting Customer 1 at position 1, Customer 2 at position 2, and no customer at position 3. The second segment, 001000000, represents Vehicle 2 visiting Customer 3 at position 1, with the remaining positions unused.

This visualization highlights the need for a contiguity treatment. If a 000 block corresponding to an unused position appears before a later visit block, such as 000100..., the encoded route contains a gap between assigned visits. This creates ambiguity in sequential travel-cost calculation and in identifying the final customer before the depot return. Therefore, a contiguity penalty is introduced in the cost Hamiltonian to consolidate unused positions toward the end of each vehicle route.

Each binary variable $x_{a,i,k}$ is mapped to a single qubit, leading to a total qubit count of $Q = |A| \times |C| \times K$, where |A| is the number of vehicles, |C| is the number of customers, and K is the maximum number of customer visits. This scaling highlights a near-term limitation of quantum routing implementations that even small routing instances require a structured allocation of qubits across vehicles, positions, and customers. The purpose of this encoding is therefore not to claim immediate scalability, but to provide a transparent representation in which transportation routing feasibility requirements can be mapped to quantum states and later preserved through mixer design.

*3.4. Cost Hamiltonian Design*

The cost Hamiltonian evaluates encoded route configurations by assigning energy values based on travel cost and, when applicable, feasibility-related penalties. In the proposed framework, it consists of three components: a travel-cost term that captures departure, intermediate travel, and return-to-depot costs; a feasibility penalty term that discourages violations of customer-visit and vehicle-position assignment requirements; and a contiguity term that prevents unused route positions from appearing before assigned customer visits. The overall cost Hamiltonian for the VRP is given in equation (13). This section constructs the VRP Hamiltonian first and then shows how it simplifies for the TSP.

From a transportation modeling perspective, the cost Hamiltonian plays the role of translating the route objective and selected feasibility treatments into an energy landscape. The travel Hamiltonian corresponds directly to the classical routing objective, while the penalty and contiguity terms represent assignment and route-continuity requirements induced by the vehicle–position–customer encoding. The Hamiltonian is designed so that lower-energy states correspond to route configurations with lower travel cost and fewer feasibility violations. In the QAOA+ formulations, the mixer further determines whether assignment-related search dynamics are restricted to feasible route states.

$$H_C^{VRP} = H_{travel}^{VRP} + \lambda_1 H_{penalty}^{VRP} + \lambda_2 H_{contiguity}^{VRP} \tag{13}$$

The first component, the travel Hamiltonian ($H_{travel}^{VRP}$), represents the total travel cost accumulated by all vehicles throughout their respective routes. As given in equation (14), it consists of three terms corresponding to departure, intermediate travel, and return cost. The first summation term measures the cost of each vehicle traveling from the depot to its first assigned customer, ensuring that the initial departure cost is included. The second term accounts for all intermediate legs between consecutive customers along the same vehicle's route, representing the total intra-route cost. Finally, the last term measures the cost of returning from the last customer to the depot. Here, $x_{a,c,k} \in \{0, 1\}$ and $y_{a,k} := \sum_{c=1}^{|C|} x_{a,c,k} \in \{0, 1\}$ where $y_{a,k} = 1$ indicates that vehicle $a$ visits exactly one customer at position $k$, and $y_{a,k} = 0$ corresponds to an unused position. For convenience in handling boundary cases, we define $y_{a,0} = 0$ and $y_{a,K+1} = 0$.

Together, these three components encode the classical VRP objective of minimizing the total travel cost of the entire fleet, as defined in the MIP formulation in *Section 3.2*. In particular, $H_{travel}^{VRP}$ serves as the quantum analog of the classical objective function by translating the route-based travel costs into phase contributions within the quantum Hamiltonian.

$$H_{travel}^{VRP} = \sum_{a=0}^{|A|-1} \left[ \sum_{k=1}^{K} \sum_{c=1}^{|C|} d_{0,c} x_{a,c,k} \left(1 - y_{a,k-1}\right) + \sum_{k=1}^{K-1} \sum_{c_1=1}^{|C|} \sum_{c_2=1}^{|C|} d_{c_1,c_2} \cdot x_{a,c_1,k} \cdot x_{a,c_2,k+1} + \sum_{k=1}^{K} \sum_{c=1}^{|C|} d_{c,0} x_{a,c,k} \left(1 - y_{a,k+1}\right) \right] \quad (14)$$

The second component, $H_{penalty}^{VRP}$, penalizes violations of assignment-related feasibility requirements in the VRP. It contains two quadratic penalty terms. The first term penalizes skipped or duplicated customer visits by assigning a nonzero penalty whenever a customer is not visited exactly once across all vehicles and route positions. The second term penalizes configurations in which more than one customer is assigned to the same vehicle-position pair. Although this requirement is expressed in a position-based assignment form rather than an arc-based flow representation, it plays a role analogous to degree and flow-consistency restrictions in the classical TSP and VRP formulations by discouraging conflicting assignments.

$$H_{penalty}^{VRP} = \sum_{c=1}^{|C|} \left(1 - \sum_{a=0}^{|A|-1} \sum_{k=1}^{K} x_{a,c,k}\right)^2 + \sum_{a=0}^{|A|-1} \sum_{k=1}^{K} \left(\sum_{c=1}^{|C|} x_{a,c,k}\right)\left(\sum_{c=1}^{|C|} x_{a,c,k} - 1\right) \quad (15)$$

The third component, the contiguity Hamiltonian ($H_{contiguity}$), is specific to the VRP formulation and addresses the empty slot issue. An empty slot occurs when no customer is assigned to a route position $k$, while a customer is assigned to the subsequent position $k+1$ (e.g. Depot - Customer 1 - not assigned - Customer 2 - Depot). This situation is problematic because $H_{travel}^{VRP}$ computes travel costs based on the sequence of assigned positions, and the presence of an unassigned intermediate position can disrupt the total cost calculation.

Unlike $H_{travel}^{VRP}$ and $H_{penalty}^{VRP}$, this term does not correspond to a standard constraint in the classical VRP formulation. In classical VRP formulations, route contiguity is implicitly guaranteed by the path-based structure of the solution and therefore does not require an explicit constraint. The contiguity penalty identifies discontinuities in the route by detecting cases where the current position $k$ of vehicle $a$ is empty while the subsequent position $k+1$ is occupied. When both conditions occur simultaneously, the product term becomes nonzero, and a penalty is added. This mechanism prevents infeasible patterns in which a route resumes after a gap, thereby enforcing continuous vehicle trips. The contiguity penalty is defined as equation (16).

$$H_{contiguity} = \sum_{a=0}^{|A|-1} \sum_{k=1}^{K-1} \left(1 - y_{a,k}\right) y_{a,k+1} \quad (16)$$

Together, the one-hot encoding and Hamiltonian components provide a quantum representation of the core routing requirements used in the classical benchmark. $H_{travel}$ encodes total travel cost, $H_{penalty}$ penalizes customer-visit and vehicle-position assignment violations when active, and $H_{contiguity}$ addresses route-continuity issues caused by unused intermediate positions. Thus, the Hamiltonian translates the classical routing objective and selected feasibility treatments into an energy landscape suitable for quantum optimization.

When the number of vehicles is reduced to one ($|A| = 1$) and the depot is fixed at both ends of the route, the formulation simplifies to the TSP, where the contiguity constraint becomes redundant and the total cost Hamiltonian reduces to

$$H_C^{TSP} = H_{travel}^{TSP} + \lambda_1 H_{penalty}^{TSP} \quad (17)$$

This structure shows that the VRP Hamiltonian generalizes the TSP case under the same optimization framework.

To examine the relative roles of constraint enforcement and search flexibility, we construct and compare three formulations: QUBO, QAOA⁺, and Hybrid QAOA⁺. The QUBO-QAOA formulation relies on penalty terms ($\lambda_1, \lambda_2 > 0$) in the cost Hamiltonian to encourage both customer-assignment feasibility and route-continuity consistency. The QAOA⁺ formulation deactivates the assignment-related feasibility penalty ($\lambda_1 = 0$) and instead uses the constraint-aware mixer to preserve assignment feasibility during the search, while retaining the contiguity term to handle unused route positions ($\lambda_2 > 0$). The Hybrid QAOA⁺ formulation combines the constraint-aware mixer with additional penalty guidance in the cost Hamiltonian. This design allows us to evaluate whether penalty-based and mixer-based mechanisms play complementary roles in guiding the search toward low-cost feasible routing configurations. Table 3 summarizes the three formulations.

**Table 3 Comparative Summary of Constraint Handling for the QUBO, QAOA⁺, and Hybrid QAOA⁺ Formulations**

| Framework | Constraint Handling Method |
|---|---|
| **QUBO-QAOA** | Assignment and route-continuity requirements encouraged through penalty terms |
| **QAOA⁺** | Assignment feasibility preserved by the mixer; route contiguity handled by the contiguity term |
| **Hybrid QAOA⁺** | Feasibility-preserving mixer combined with additional penalty guidance |

*3.5. Constraint-Preserving Mixer Hamiltonian Design*

The proposed mixer is designed to play the role of a feasibility-preserving routing neighborhood operator within the quantum circuit. Under the vehicle–position–customer encoding, a route configuration can be viewed as a stack of customer-position assignment tables, one for each vehicle. The mixer operates through column-wise swaps, which exchange assignment columns within or across vehicle-specific tables while preserving the one-hot and customer-visit uniqueness constraints. Here, a column refers to a route-position column in a vehicle-specific customer-position assignment table. In this study, two types of column-wise swaps are used: intra-vehicle column-wise swaps, which reorder customer visits within a vehicle route, and inter-vehicle column-wise swaps, which reassign customers across vehicles while preserving assignment feasibility. Figure 3 illustrates these two types of column-wise swaps.

An intra-vehicle swap exchanges two customer assignments within the same vehicle route. In Figure 3, the customers assigned to position 1 and position 2 of Vehicle 1 are swapped, which changes the visit sequence without changing the set of customers served by that vehicle. An inter-vehicle swap exchanges customer assignments across different vehicles. In Figure 3, a customer assigned to position 2 of Vehicle 1 is exchanged with a customer assigned to position 2 of Vehicle 2, which reallocates customers across vehicles while preserving the assignment structure. Both moves preserve the at-most-one-customer-per-position condition and the customer-visit uniqueness condition.

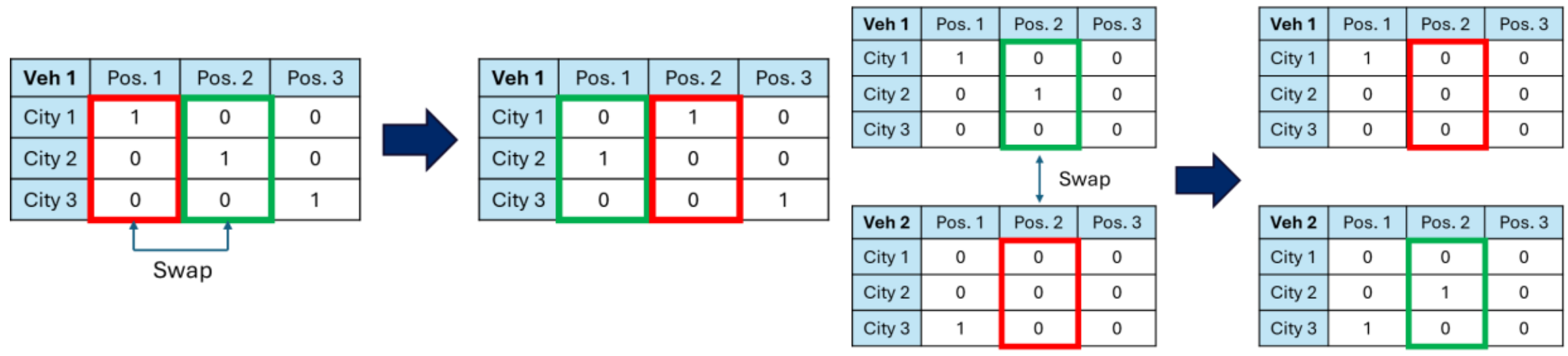

**Figure 3 Illustration of the two components of the constraint-preserving mixer: intra-vehicle swap (left) and inter-vehicle swap (right).**

The mixer Hamiltonian is composed of two complementary components: a jumper mixer and a search mixer. Such hierarchical exploration strategy closely mirrors the diversification–intensification paradigm widely adopted in classical metaheuristic algorithms, such as Tabu Search (Glover et al., 2007) and Variable Neighborhood Search (Hansen and Mladenovi, 2001). In implementation, the cost, jumper, and search operators are applied sequentially within each layer.

The jumper mixer plays a role of global diversification by enabling broader transitions among feasible routing configurations. When applied with a fixed rotation angle, the XY-type coupling in equation (18) induces full column-wise swaps between feasible route configurations (e.g., $|01\rangle \leftrightarrow |10\rangle$). In transportation terms, these full swaps correspond to larger route modifications, such as resequencing customer visits within a vehicle route or reallocating customers across vehicles. In equation (18), the first summation corresponds to inter-vehicle column-wise exchanges, while the second summation corresponds to intra-vehicle column-wise exchanges.

$$H_{jump} = \sum_{a_1<a_2} \sum_{c} (X_{a_1,c,k} X_{a_2,c,k} + Y_{a_1,c,k} Y_{a_2,c,k}) + \sum_{a} \sum_{c_1<c_2} (X_{a,c_1,k} X_{a,c_2,k} + Y_{a,c_1,k} Y_{a,c_2,k}) \quad (18)$$

During evolution, the jumper mixer is applied with a fixed rotation angle of $\pi/4$ for the deterministic jump between different routing configurations. The unitary operator generated by $H_{jump}$ is given in equation (19). Here, $H_{jump}$ denotes the active jump Hamiltonian at a given layer; in practice, intra-vehicle and inter-vehicle jump couplings are applied alternately across odd and even layers.

$$U_{jump} = \exp(-i(\pi/4)H_{jump}) \quad (19)$$

Relying only on full-swap transitions would limit the circuit's ability to form superpositions over neighboring feasible routes. The search mixer is therefore introduced to support local refinement. It creates superpositions over feasible candidate states in the neighborhood of the current route configuration, allowing the circuit to explore local variations while preserving assignment-related feasibility. The formulation for the search mixer is given in equation (20).

$$H_{search} = \sum_{(i,j)\in\mathcal{N}_{local}} (X_i X_j + Y_i Y_j), \quad (20)$$

Here, $\mathcal{N}_{local}$ denotes the set of feasible intra- and inter-vehicle column-wise swaps that preserve one-hot and customer-visit uniqueness constraints. The unitary operator generated by $H_{search}$ is expressed as equation (21). This operator is applied with a smaller, tunable rotation angle $\beta/2$, which performs partial swaps between feasible configurations, splitting amplitude between $|01\rangle$ and $|10\rangle$ rather than fully exchanging them. As a result, the search mixer generates superpositions of valid routing states, allowing local refinement around promising feasible configurations discovered by the jump layer.

$$U_{search} = \exp(-i(\beta/2)H_{search}) \quad (21)$$

With the mixer Hamiltonians and their corresponding unitary operators formally defined, this forms the basis for the following feasibility guarantee, formalized in Proposition 1. In our discussion in this section, a feasible configuration is defined as a binary assignment $x_{a,c,k}$ satisfying the one-hot constraint $\sum_{c=1}^{|C|} x_{a,c,k} = 1 \forall\, a, k$, together with the permutation constraint $\sum_{a=0}^{|A|-1} \sum_{k=1}^{K} x_{a,c,k} = 1$ for each customer $c \in C$ (excluding the depot).

To formalize the key property of the column-wise swap mixer, let

$$\Omega_F = \left\{ x_{a,c,k} \in \{0,1\} \middle| \sum_{c=1}^{|C|} x_{a,c,k} = 1 \forall\, a, k\,, \sum_{a=0}^{|A|-1} \sum_{k=1}^{K} x_{a,c,k} = 1\ \forall c \in C \right\}$$

denote the set of feasible routing assignments satisfying the one-hot and permutation constraints. Define the feasible Hilbert space as $\mathcal{H}_F \coloneqq span\{|x\rangle : x \in \Omega_F\}$ i.e., the linear span of computational basis states corresponding to feasible assignments. For the search mixer, we define $N_{\text{local}}$as the set of index pairs $(i, j)$corresponding to assignment variables such that exchanging the two associated entries in any feasible assignment yields another feasible assignment; that is, each swap in $N_{\text{local}}$preserves both the one-hot and permutation constraints.

---

**Proposition 1. Feasibility Preservation of the Column-wise Swap Mixer**

Let the column-wise swap mixer $U_M = U_{search}(\beta)U_{jump}\left(\frac{\pi}{4}\right)$ with $U_{search}$ and $U_{jump}$ as defined in (18)-(21). Then $U_M$ preserves feasibility, i.e., preserves $\mathcal{H}_F$ under its unitary evolution: if $|\psi\rangle \in \mathcal{H}_F$, then $U_M|\psi\rangle \in \mathcal{H}_F$.

---

**Proof.** It suffices to show that both $U_{jump}$ and $U_{search}$ map $\mathcal{H}_F$ to itself. Since $\mathcal{H}_F$ is closed under composition, invariance under $U_M$ follows immediately.

**(i) Preservation under $U_{jump}$.**

Each term in $H_{\text{jump}}$ has the form

$$g_{pq} = X_p X_q + Y_p Y_q,$$

where $p$ and $q$correspond to two assignment variables $x_{a_1,c,k}$ and $x_{a_2,c,k}$ (inter-vehicle exchange) or $x_{a,c_1,k}$ and $x_{a,c_2,k}$ (intra-vehicle exchange), all at the same time position index $k$.

The operator $g_{pq}$ acts on the two-qubit computational basis as

$$g_{pq}|00\rangle = 0, \qquad g_{pq}|11\rangle = 0,$$
$$g_{pq}|01\rangle = 2|10\rangle, \qquad g_{pq}|10\rangle = 2|01\rangle.$$

Fix any feasible computational basis state $|x\rangle = \mathcal{H}_F$. Then

$$g_{pq}|x\rangle \in \text{span}\{|x\rangle, |x^{(p\leftrightarrow q)}\rangle\}$$

where $|x^{(p\leftrightarrow q)}\rangle$ denotes the assignment obtained by swapping the binary values at positions $p$ and $q$.

Because each $g_{pq}$ corresponds to an exchange of assignment variables at the same position $k$, such swaps preserve the one-hot constraint $\sum_{c=1}^{|C|} x_{a,c,k} = 1 \forall\, a, k$ and the customer-count constraint $\sum_{a=0}^{|A|-1} \sum_{k=1}^{K} x_{a,c,k} = 1\ \forall c \in C$, since no customer index is created or removed – only reassigned.

Hence, $g_{pq}|x\rangle \in H_F$. Because $H_{\text{jump}} = \sum g_{pq}$ is a sum of such operators, it follows that $H_{\text{jump}}\mathcal{H}_F \subseteq \mathcal{H}_F$.

Finally, since $U_{jump} = \exp\left(-i\left(\frac{\pi}{4}\right)H_{jump}\right)$ admits a power series expansion in $H_{\text{jump}}$ we conclude that $U_{\text{jump}}(\pi/4)\mathcal{H}_F \subseteq \mathcal{H}_F$. Thus $U_{\text{jump}}$ preserves feasibility.

It follows that $U_{jump}$ maps $\mathcal{H}_F$ to itself.

**(ii) Preservation under $U_{search}$.**

By definition, $\mathcal{N}_{local}$ contains only index pairs $(i, j)$ such that exchanging the corresponding assignment variables yields another feasible assignment. Each term in $H_{search}$ therefore generates partial swaps between two assignment variables whose exchange preserves feasibility.

Fix any feasible computational basis state $|x\rangle \in \mathcal{H}_F$. Consider a single term $h_{ij} = X_i X_j + Y_i Y_j$ with $(i, j) \in N_{\text{local}}$. As shown in part (i), $h_{ij}$ has nonzero action only on basis components where the qubits $(i, j)$ are in the states $|01\rangle$ or $|10\rangle$; moreover,

$$h_{ij}|x\rangle \in \text{span}\{|x\rangle, |x^{(i\leftrightarrow j)}\rangle\},$$

where $|x^{(i\leftrightarrow j)}\rangle$ denotes the basis state obtained by swapping the binary values at indices $i$ and $j$ (and if $(i, j)$ acts on $|00\rangle$ or $|11\rangle$, then $h_{ij}|x\rangle = 0$). By construction of $N_{\text{local}}$, whenever a swap $(i, j)$ is active (i.e., corresponds to a $|01\rangle$ or $|10\rangle$ pair), the swapped assignment $x^{(i\leftrightarrow j)}$ is also feasible. Hence $h_{ij} \mid x\rangle \in \mathcal{H}_F$.

Since $H_{\text{search}} = \sum_{(i,j)\in N_{\text{local}}} h_{ij}$ is a sum of operators each mapping feasible basis states into $\mathcal{H}_F$, it follows that

$$H_{\text{search}}\mathcal{H}_F \subseteq \mathcal{H}_F.$$

---

Finally, because $U_{\text{search}}(\beta) = \exp(-i(\beta/2)H_{search})$ can be written as a power series in $H_{\text{search}}$,

$$U_{\text{search}}(\beta) = \sum_{m=0}^{\infty} \frac{1}{m!}\left(-i\frac{\beta}{2}H_{\text{search}}\right)^{m},$$

and $H_{\text{search}}\mathcal{H}_F \subseteq \mathcal{H}_F$ for all $m$, we conclude that $U_{\text{search}}(\beta)\mathcal{H}_F \subseteq \mathcal{H}_F$. Therefore $U_{\text{search}}$ preserves feasibility.

Within each QAOA+ layer, intra- and inter-vehicle exchanges in the jumper mixer are alternated across odd and even layers, while the search mixer includes both types of local couplings at every layer. This alternating design balances global diversification and local intensification across layers. Accordingly, each QAOA+ layer is defined as equation (22). The algorithm to generate the whole layer is summarized in Algorithm 1.

$$U_{layer} = U_{search}(\beta)U_{jump}(\pi/4)U_{cost}(\gamma) \tag{22}$$

Algorithm 1 QAOA+ Layer Construction with Column-wise Constraint-Preserving Mixers

Input:

Feasible routing state encoded in one-hot form
Parameters: $\gamma_l$(cost), $\beta_l$(search)
Fixed rotation angle $\pi/4$ (jumper)

for layer $l = 1, 2, \dots, p$ do
1. Apply cost unitary:
   $U_{cost}(\gamma_l)$.
2. Apply jumper mixer
   if problem is TSP then
      apply intra-vehicle jump only
   else
      if $l$ is odd then
         activate intra-vehicle swap jump
      else
         activate inter-vehicle swap jump
      end if
   end if
   $U_{jump} = exp(-i(\pi/4)H_{jump})$
3. Apply search mixer
   restrict $XX + YY$ couplings to locally feasible qubit pairs
   $U_{search} = exp(-i(\beta_l/2)H_{search})$
end for

Output: Superposition of feasible routing configurations

*3.6. Solution Extraction and Evaluation*

The output of a quantum circuit is inherently probabilistic due to quantum measurement postulates. After the parameterized cost and mixer unitaries are applied, the quantum state collapses into a computational basis state upon measurement, producing a distribution of bitstrings. Each bitstring encodes a candidate route configuration in the TSP or VRP formulation, and its measurement frequency (probability) estimates the sampling probability assigned to that route configuration by the optimized circuit.

In the context of transportation systems, this probability distribution can be interpreted as an algorithmic distribution over candidate routing solutions. For example, if the optimal route (minimum travel cost) appears with a measurement probability of 25%, it implies that under the current QAOA parameters, this route is sampled with 25% probability. Rather than representing behavioral route choice, this distribution is used here to evaluate how each formulation allocates probability mass across feasible, infeasible, low-cost, and near-optimal route configurations. From a decision-making perspective, this provides both a primary recommendation, based on the most probable route, and an indication of probability concentration across competing feasible routes. A sharply peaked distribution suggests strong concentration around a small number of feasible routing solutions, whereas a broader one indicates multiple competing near-optimal solutions.

To systematically evaluate the performance of the quantum optimization framework, each sampled bitstring is post-processed and analyzed according to three complementary criteria: feasibility, travel cost, and probability of occurrence. First, a feasibility check ensures that the decoded bitstring satisfies all problem constraints, such as one-customer-per-position, each customer being visited exactly once, and route contiguity. Although the QAOA$^{+}$ mixer is designed to preserve assignment-related feasibility, full decoded route feasibility also depends on the contiguity treatment and post-measurement decoding criteria. Infeasible bitstrings are excluded from the performance metrics but remain informative about the exploratory behavior of the mixer Hamiltonian, indicating whether the algorithm is maintaining the feasible search space throughout the optimization process.

Second, for all feasible routes, the total cost is computed classically using the given cost matrices. The expected value of the cost Hamiltonian estimated during QAOA execution corresponds to the energy expectation of this objective, whereas the decoded route with the lowest computed cost represents the empirically optimal solution found in the sampling stage.

Third, the probability of occurrence of each feasible route is analyzed to construct a distribution over objective value. By aggregating measurement probabilities of routes with similar travel costs, the resulting histogram illustrates how the quantum circuit concentrates probability mass around low-cost solutions. This distributional view enables quantitative comparison among different formulations, QUBO-QAOA, QAOA$^{+}$, and Hybrid QAOA$^{+}$, in terms of their probabilistic focus on solutions.

*3.7. Experimental Design and Benchmarking Framework*

The input data for all experiments were derived from predefined TSP/VRP vehicle-dependent travel cost matrices, as presented in Table 4. These matrices determine the base cost scale $c_{\max} = \max_{i \neq j} c_{ij}$, which is used for penalty normalization and parameter calibration throughout the experiments. The problem instances considered in this study are intentionally small, since the primary objective of the experiments is not to demonstrate large-scale computational performance, but to evaluate the structural behavior and feasibility-preserving properties of the proposed mixer design under controlled settings.

**Table 4 TSP / VRP Vehicle-dependent Travel Cost Matrices**

| Parameters | Value |
|---|---|
| **Vehicle 1** | [0 2 9 10], [2 0 6 4], [9 6 0 8], [10 4 8 0] |
| **Vehicle 2** | [0 3 7 11], [3 0 5 6], [7 5 0 9], [11 6 9 0] |

All experimental parameters were chosen based on theoretical and empirical guidelines established in recent QAOA studies, balancing expressive power, statistical reliability, and constraint stability for the number of layers ($p$), number of measurement shots, and penalty magnitudes. The number of layers was limited to small values to avoid excessive circuit depth, which is known to lead to diminishing performance gains and rapid error accumulation beyond a saturation point (Niu et al., 2019). Based on these considerations, we set $p \in \{2,3\}$, which allows both intra-vehicle and inter-vehicle swaps to be activated while keeping circuit depth within a controlled NISQ-compatible range for the problem sizes studied.

The number of measurement shots was chosen to ensure stable probability estimates while maintaining feasible runtime. As the precision of estimated outcome probabilities improve with the number of circuit repetitions, several thousand shots are commonly used to reduce sampling variability in NISQ-era variational experiments (Seksaria and Prabhakar, 2025). Following these established practices, we used 4096 shots per circuit evaluation, a scale widely adopted in NISQ-era variational experiments to balance statistical reliability and computational cost.

Penalty coefficients in the cost Hamiltonian were scaled relative to the maximum single-move cost $c_{max}$. Prior studies suggest that penalty weight must be sufficiently large to discourage infeasible solutions relative to feasible alternatives (Ayodele, 2022), while overly large penalties can dominate the energy landscape and hinder convergence by suppressing meaningful exploration (Glover et al., 2022). To maintain a balance between feasibility enforcement and search flexibility, we normalized penalty weights by $c_{max}$, ensuring that constraint violations are discouraged without overwhelming the travel cost term. The contiguity penalty ($\lambda_2$) was calibrated based on the mean value of the cost matrix with $\bar{c} \approx 6.8$, yielding $\lambda_2 = 15$. This choice softly encourages route continuity while avoiding excessive distortion of the objective landscape. The complete set of selected experimental parameters is summarized in Table 5.

**Table 5 Parameters Used for QUBO-QAOA, QAOA$^+$, and Hybrid QAOA$^+$ implementations**

| Parameters | Value |
|---|---|
| **Circuit depth $(p)$** | 2, 3 |
| **Number of shots $(n)$** | 4,096 |
| **Constraint penalty $(\lambda_1)$** | QUBO: 60/Hybrid QAOA$^+$: 5/QAOA$^+$: 0 |
| **Contiguity penalty $(\lambda_2)$** | 15 |

To provide a reference for solution optimality, we additionally computed the classical exact optima for both TSP and VRP instances using mixed-integer programming formulations implemented in Gurobi optimizer. The TSP baseline follows the Dantzig-Fulkerson-Johnson (DFJ) formulation described in equations (1)-(5), which minimizes the total travel cost while enforcing one-visit-per-city and eliminating subtours through subset-based constraints. For the VRP instances, this formulation was extended by incorporating depot and fleet constraints as described in equations (6)–(12), allowing multiple vehicle routes while maintaining mandatory depot return conditions. These exact solutions provide the optimal cost $C^*$ for each instance and serve as benchmarks for evaluating quantum results.

For quantitative comparison between quantum and classical approaches under identical cost definitions, the best feasible sampled route cost $C_{best}$ was normalized by the classical optimum $C^*$ to compute the relative cost ratio as:

$$Relative\ cost\ ratio = \frac{C_{best}}{C^*}. \quad (23)$$

For this minimization problem, a value closer to 1 indicates better agreement with the classical optimum, while larger values indicate a larger optimality gap. This metric enables consistent comparison across different formulations, including QAOA$^+$, Hybrid QAOA$^+$, and QUBO-QAOA.

## 4. Comparative Results and Discussion

This section presents a comparative performance analysis of three quantum algorithm variants: QAOA$^+$, Hybrid QAOA$^+$, and QUBO-QAOA. These algorithms were applied to small scale TSP and VRP instances. Simulations were performed using the Qiskit Aer simulator with the COBYLA optimizer and 4096 measurement shots per evaluation. Algorithm performance was evaluated at QAOA depths of $p = 2$ and $p = 3$. The evaluation focuses on three key metrics: (i) the ability to identify optimal or feasible solutions, (ii) the relative cost ratio, (iii) the probability of sampling near-optimal solutions (within 5% of the optimum), in addition to convergence behavior. The purpose of these experiments is not to establish large-scale performance advantages, but to isolate how different constraint-handling architectures influence search behavior within the feasible solution space. The results are therefore discussed not only in terms of solution quality, but primarily in terms of how different mixer designs shape the feasible search landscape.

Although the numerical experiments compare quantum formulations rather than classical routing algorithms, the results can be interpreted relative to classical feasibility-preserving search principles used in routing. In both exact and heuristic routing methods, feasibility is maintained by controlling admissible states and transitions—through formulation structure and route generation in exact methods, and through feasibility-preserving or feasibility-restoring neighborhood operators (e.g., swap, relocate, insertion, and related moves) in metaheuristics. These mechanisms update one incumbent solution through sequential transitions within (or near) the feasible set.

The proposed QAOA$^+$ and Hybrid QAOA$^+$ formulations follow the same methodological principle of structuring the search to avoid unnecessary exploration of invalid route configurations but implement it through a different mechanism. Specifically, the constraint-aware mixer restricts quantum evolution to assignment-feasible route states while probability amplitude is redistributed across multiple configurations before measurement. For example, the intra- and inter-column swap operations implemented in the mixer mirror classical swap/relocate-style neighborhood transitions while guaranteeing one-hot and permutation feasibility. This provides a quantum analogue of feasibility-preserving neighborhood search, where feasibility is enforced at the operator level rather than through post-hoc repair or penalty calibration.

Accordingly, the purpose of the experiments is not to claim that the proposed quantum formulations outperform mature classical routing algorithms, but to isolate how different quantum constraint-handling architectures behave when feasibility-preserving search logic is embedded into a quantum operator design. In this sense, the results should be read as evidence on how mixer-based feasibility preservation reshapes quantum routing search behavior, including feasible-solution sampling, convergence patterns, and probability concentration over low-cost feasible routes.

*4.1 Solution Quality and Optimality*

First, classical optimal solutions were computed as benchmarks: 23.0 for the $|N| = 4$ TSP instance and 30.0 for the $|N| = 4, K = 3, |A| = 2$ VRP instance. The performance summary, including the optimality ratio (best QAOA result cost/classical optimum), is presented in Table 6.

Table 6 summarizes the performance of each method in terms of best achieved cost and the probability of sampling near-optimal feasible solutions. For the TSP, while the Hybrid reaches the optimal cost (23.0) and consistently concentrates a substantial probability mass on optimal solutions (approximately 44.7% at $p = 2$ and 41.6% at $p = 3$), standard QAOA$^+$ and QUBO fail to reach the optimal cost, often resulting in suboptimal or infeasible states. QAOA$^+$ achieved a near-optimal sample only at $p = 2$ with negligible probability and becoming unreliable at $p = 3$, while the QUBO formulation fails to find the optimal or near-optimal solution. For the TSP, penalty-free QAOA$^+$ avoids some infeasible exploration but does not substantially concentrate probability near the optimum.

This observation suggests that, in small permutation-based problems, feasibility preservation alone may be insufficient to significantly bias the sampling distribution toward high-quality solutions without additional landscape shaping. In particular, the TSP feasible subspace is already highly symmetric and low-dimensional, such that restricting the dynamics to feasible tours does not sufficiently break degeneracies or amplify energy gradients toward the optimum.

Interestingly, the VRP results exhibit a different trend. In this case, the constraint-aware QAOA$^+$ achieves consistently non-negligible near-optimal sampling probabilities, indicating that explicit feasibility preservation plays a more dominant role than additional penalty-based shaping. This contrast between TSP and VRP highlights that the effectiveness of hybridization is problem-structure dependent and motivates a closer examination of the underlying convergence behavior and probability distributions.

**Table 6 Performance Summary by Method and Depth**

| Problem | Method | $p$ | Best QAOA$^+$ Cost | Classical Opt. | Cost Ratio | P (within 5% of the optimum) |
|---|---|---|---|---|---|---|
| TSP | QAOA$^+$ | 2 | 26.0 | 23.0 | 1.1 | 0.002 |
| | | 3 | - | 23.0 | Failed* | 0.011 |
| | Hybrid QAOA$^+$ | 2 | 23.0 | 23.0 | 1.0 | 0.447 |
| | | 3 | 23.0 | 23.0 | 1.0 | 0.416 |
| | QUBO-QAOA | 2 | - | 23.0 | Failed | 0.036 |
| | | 3 | - | 23.0 | Failed | 0.003 |
| VRP | QAOA$^+$ | 2 | 30.0 | 30.0 | 1.0 | 0.222 |
| | | 3 | 30.0 | 30.0 | 1.0 | 0.201 |
| | Hybrid QAOA$^+$ | 2 | 30.0 | 30.0 | 1.0 | 0.253 |
| | | 3 | 33.0 | 30.0 | 1.1 | 0.120 |
| | QUBO-QAOA | 2 | - | 30.0 | Failed | 0.014 |
| | | 3 | - | 30.0 | Failed | 0.003 |

* Indicates that the run did not produce the finite best feasible cost for defining the optimality ratio, although near-optimal samples may still appear with low probability

*4.2 Convergence Behavior*

The differences in solution quality observed in Table 6 can be further understood by examining the convergence behavior of each method, as shown in Figure 4 and Figure 5. These figures report normalized energy improvement across optimizer iteration of TSP and VRP, respectively, where 1 represents the starting energy and 0 represents the final converged energy.

For the TSP, Figure 4 shows that the Hybrid QAOA$^+$ (blue line) converges faster and reaches a low-energy regime within fewer optimizer iterations than other methods. The QAOA$^+$ (orange line) converges more slowly and to a higher relative energy. While the QUBO approach (green lines) eventually reaches a converged state at 0, it exhibits the slowest progress among all methods. Most importantly, as indicated by the results in Table 6, this convergence does not correspond to the discovery of an optimal solution, instead, it represents a termination in infeasible or suboptimal regions. This indicates that the Hybrid variant's landscape is the most effective not just in speed, but in guiding the search toward the known optimal solution for this instance while avoiding the suboptimal traps encountered by the other formulations.

This pattern, however, does not directly carry over to the VRP case. As shown in Figure 5, the convergence behavior exhibits a different trend. While the QUBO (solid green line) appears to converge fastest, this apparent convergence corresponds to premature trapping in an infeasible, high-energy state, as indicated by the results in Table 6. In contrast, both the standard QAOA$^+$ (orange lines) and Hybrid QAOA$^+$ (blue lines) exhibit more gradual convergence, reflecting continued exploration within the feasible solution space. This slower convergence behavior is associated with improved solution quality. As Table 6 confirms, the standard QAOA$^+$ runs ($p = 2$ and $p = 3$) converge to the true global optimum (cost 30.0), while the Hybrid QAOA$^+$ ($p = 3$) converges to a valid but suboptimal solution (cost 33.0). These results indicate that, unlike the TSP case where moderate intensification is beneficial, broader exploration plays a more critical role in navigating the more complex feasible landscape of the VRP.

Taken together, the convergence results highlight that the effectiveness of different constraint-handling strategies depends strongly on the structure of the underlying routing problem. The Hybrid QAOA$^+$ performs well in the simpler TSP setting, where the feasible space is relatively compact, while the standard feasibility-preserving QAOA$^+$ is more effective in the VRP, where avoiding premature convergence and maintaining exploration diversity are more important.

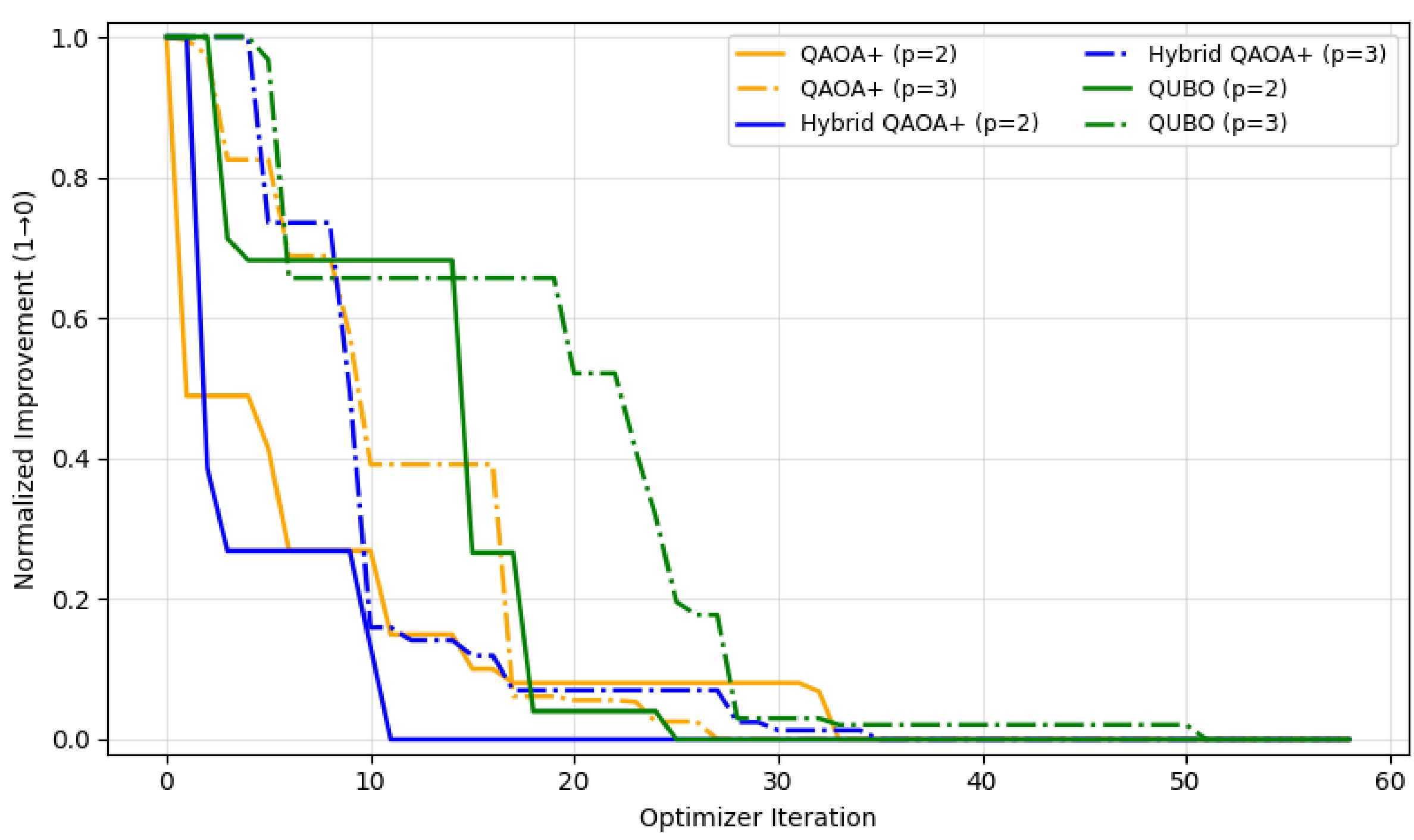

**Figure 4 Convergence behavior of the QUBO-based QAOA, penalty-free QAOA$^+$, and Hybrid QAOA$^+$ formulations of the TSP**

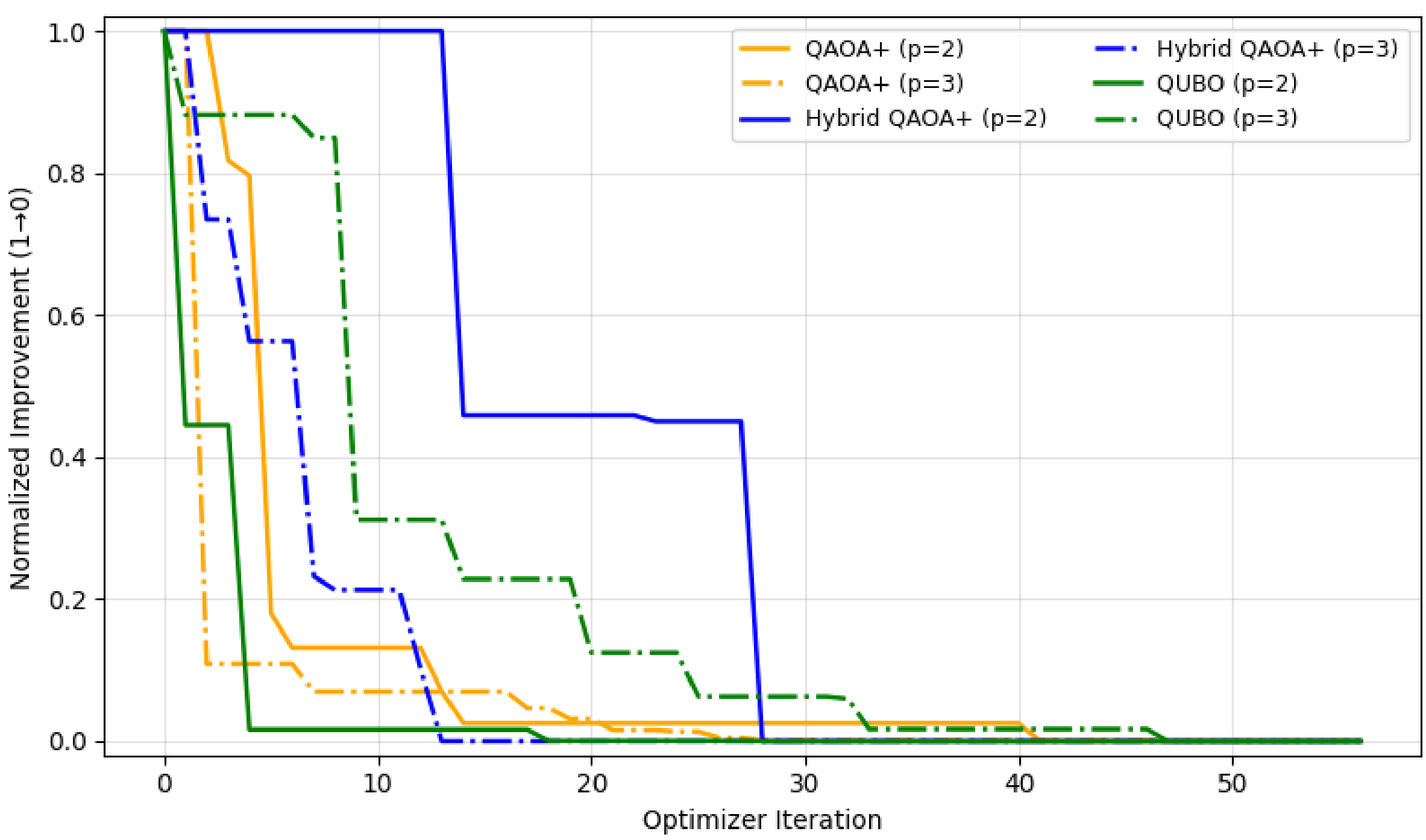

**Figure 5 Convergence behavior of the QUBO-based QAOA, penalty-free QAOA$^+$, and Hybrid QAOA$^+$ formulations of the VRP**

*4.3 Probability Distribution*

The convergence behavior observed in Section 4.2 is further reflected in the probability distributions over objective values at $p = 2$, as shown in Figure 6 and Figure 7. These figures show the probability (y-axis) of observing different objective values (x-axis), distinguishing between feasible (darker bars) and infeasible (lighter, hatched bars) solutions. The QUBO distribution (green) is widely spread across a vast range of both feasible and infeasible states, many with high objective values. This illustrates the inefficiency of penalty-based QUBO-QAOA exploration, which fails to confine the search to the valid subspace and therefore allocates probability mass to infeasible, high-energy states. For the TSP, this explains why QUBO yields negligible (<4%) near-optimal sampling probabilities in Table 6.

This performance gap for the TSP is significant. The QAOA$^+$ (orange bars in Figure 6), while preserving assignment-related constraints, explores the state space more broadly, leading to a less concentrated probability mass near the optimum (Ratio 1.1 at $p = 2$, Failed at $p = 3$). Conversely, the Hybrid QAOA$^+$ approach (blue bars in Figure 6) adds a small soft-constraint penalty. For the simple, single-vehicle permutation structure of the TSP, this penalty acts as an effective intensification mechanism. It successfully focuses the search on the high-quality feasible subspace without sacrificing necessary exploration, as evidenced by its 1.0 optimality ratio and high near-optimal sampling probability (approx. 44.7% at $p = 2$ and 41.6% at $p = 3$).

The VRP results reveal a different performance pattern between the two QAOA$^+$ variants: the standard penalty-free QAOA$^+$ remains robust, whereas the Hybrid QAOA$^+$ becomes more prone to local trapping as depth increases. Similarly, the QUBO-based QAOA struggled to produce feasible solutions or converge effectively, resulting in no finite best feasible sampled cost and very low near-optimal sampling probabilities. The probability distribution for VRP at $p = 2$ (Figure 7) again shows the QUBO results (green) widely dispersed across high-energy and predominantly infeasible states. The penalty-free QAOA$^+$ (orange bars in Figure 7) performed well, successfully finding the optimal solution (cost 30.0, ratio 1.0) for both $p = 2$ and $p = 3$, with substantial near-optimal sampling probabilities (22.2% for $p = 2$, 20.1% for $p = 3$). Its probability distribution shows clear peaks in the low-energy, feasible region. The QAOA$^+$ (Hybrid) method (blue bars in Figure 7) also found the optimum at $p = 2$ (ratio 1.0, sampling probability 25.3%), exhibiting a similar concentration of probability on feasible, low-energy states. However, it converged to a suboptimal feasible solution (cost 33.0, ratio 1.1) at $p = 3$, with a reduced sampling probability (12.1%).

This performance reversal, where the standard QAOA$^+$ outperforms the hybrid model on the more complex VRP, appears to arise from the interaction between the problem's multi-vehicle structure and the additional penalty guidance in the Hybrid QAOA$^+$ formulation. The added penalty term in the hybrid objective sharpens the landscape inside the feasible set, which can accelerate convergence when the feasible region is compact (as in the TSP). For the VRP, however, the feasible set decomposes into multiple structurally distinct allocations across vehicles, and the same intensification can over-concentrate probability around a limited partition, reducing effective inter-vehicle exploration. Consequently, it restricts the inter-vehicle search transitions necessary to explore the broader feasible space, leading to a suboptimal convergence at cost 33.0 (ratio = 1.1, P = 12.1%).

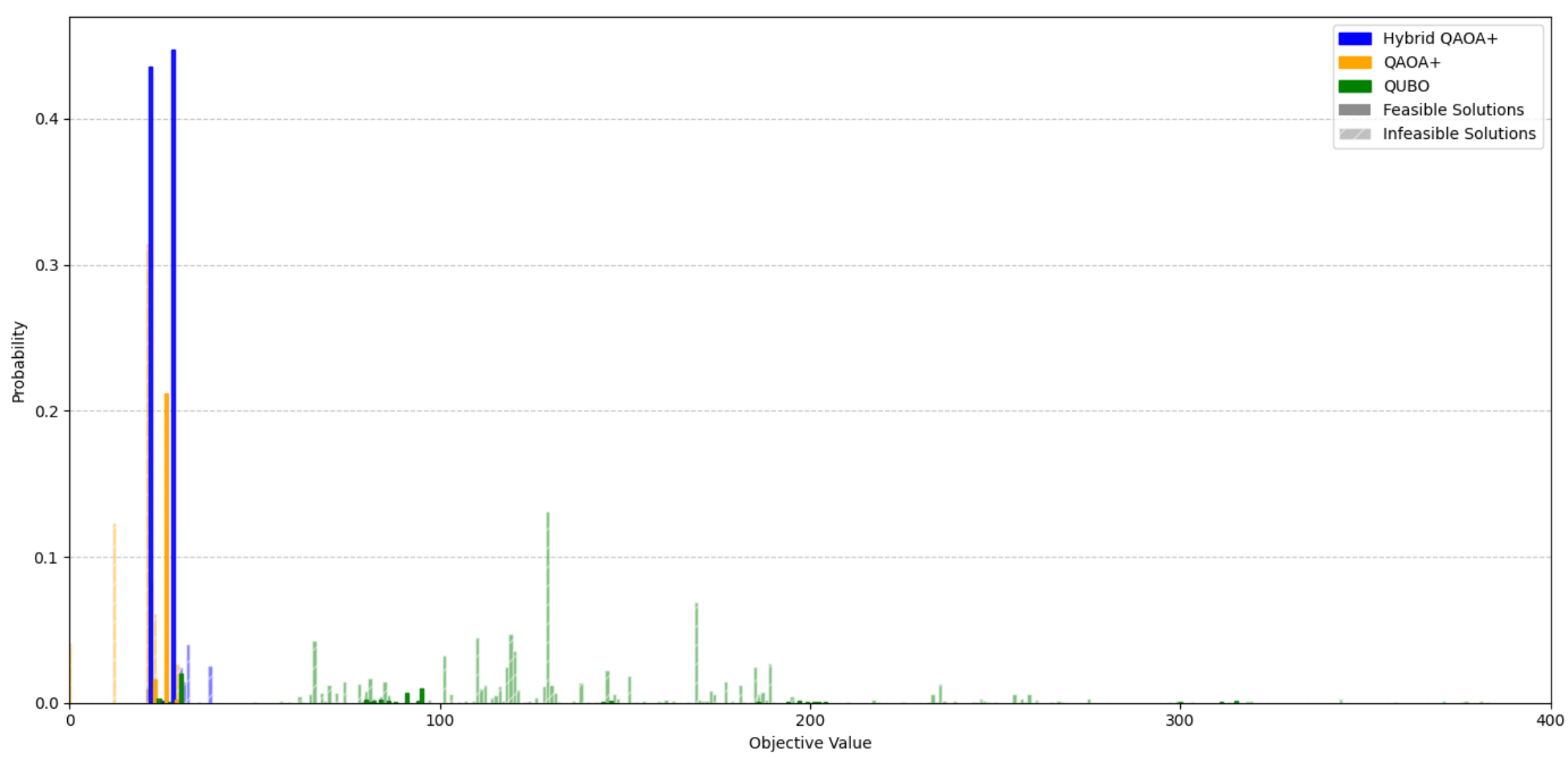

**Figure 6 Measurement probability distribution for the TSP ($p = 2$), comparing QUBO-based QAOA, penalty-free QAOA$^+$, and Hybrid QAOA$^+$ formulations.**

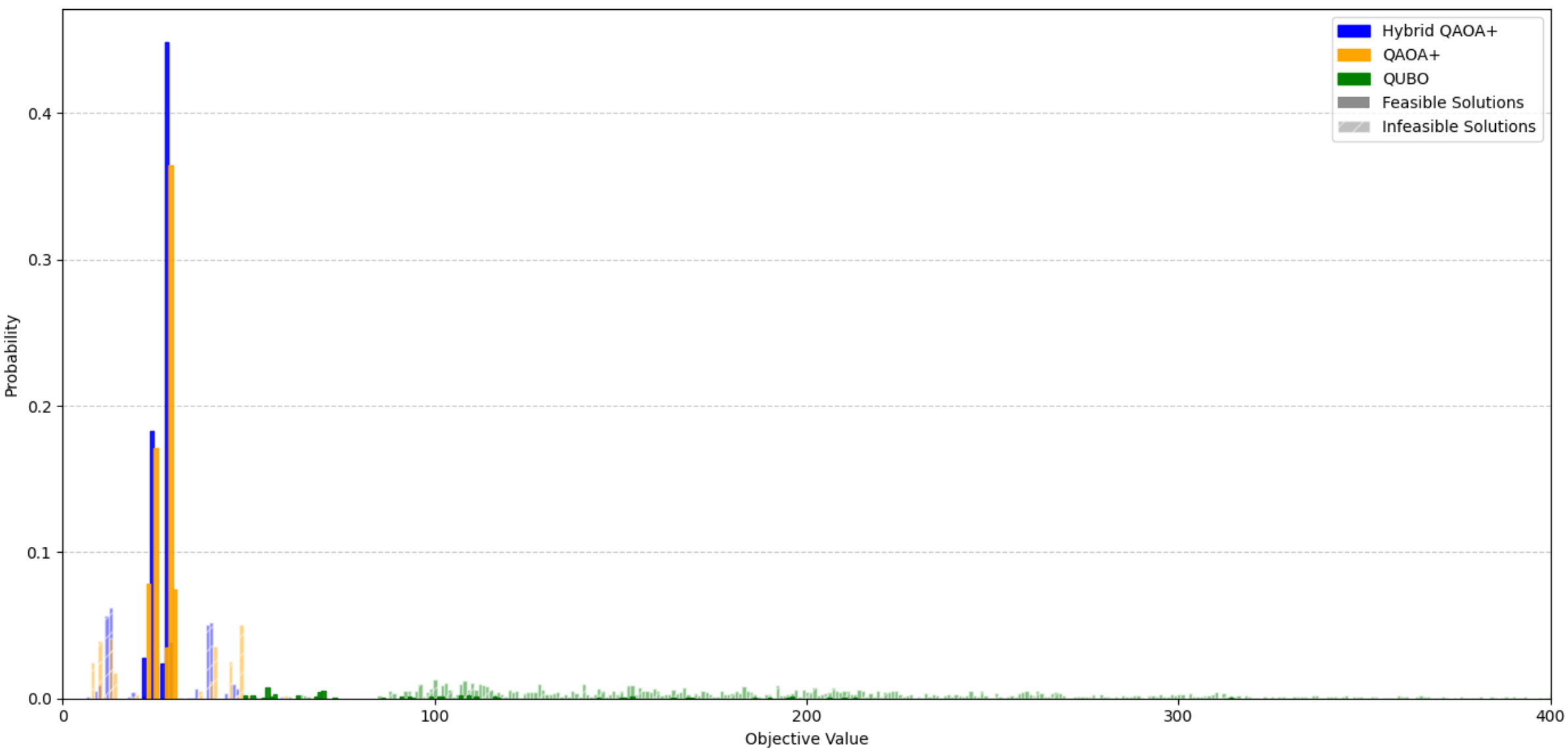

**Figure 7 Measurement probability distribution for the VRP ($p = 2$), comparing QUBO-based QAOA, penalty-free QAOA$^+$, and Hybrid QAOA$^+$ formulations.**

### *4.4 Discussion and Implications for Near-Term Quantum Hardware*

The comparative results indicate that in the controlled small instances examined here, feasibility-preserving QAOA$^+$ formulations generally produce more useful feasible-route samples than the penalty-based QUBO-QAOA formulation. While deeper circuits (p = 3) generally improved convergence towards lower final energy values, increased depth does not uniformly translate into improved feasible solution quality. The relative performance of Hybrid and penalty-free QAOA$^+$ variants further depends on problem structure: moderate intensification benefits the compact feasible landscape of the TSP, whereas broader exploration is more effective for the multi-vehicle VRP.

These findings underscore that embedding routing constraints directly into the quantum evolution through mixer design is central to reliable performance. The effectiveness of penalty-based shaping appears to depend on the topology of the feasible space, suggesting that constraint-aware operator design should be tailored to problem structure. This reinforces the need for topology-sensitive quantum optimization strategies in constrained transportation routing applications.

Beyond these problem-level observations, the results carry important implications for near-term quantum hardware. Current quantum processors operate under the constraints of the Noisy Intermediate-Scale Quantum (NISQ) era, where both the number of available qubits and the achievable circuit depth are limited. In the routing formulation used in this study, the number of required qubits scales as $|A| \times |C| \times K$, growing rapidly with the number of vehicles, customers, and route positions. Even modest increases in problem size therefore led to substantial qubit requirements. At the same time, deeper QAOA circuits (larger $p$), while potentially improving expressiveness, introduce greater sensitivity to noise and increased computational overhead.

Under these hardware constraints, the distinction between penalty-based and feasibility-preserving formulations becomes particularly important. Penalty-based QUBO approaches explore a large infeasible portion of the solution space, allocating quantum resources to configurations that cannot correspond to valid routes. This inefficient exploration effectively increases the depth and sampling effort required to identify feasible high-quality solutions. In contrast, feasibility-preserving designs reduce unnecessary exploration of invalid assignment configurations and improve efficiency in sampling under limited hardware resources. These considerations highlight that scalability challenges in quantum routing are not only driven by problem size, but also by how effectively constraint structure is embedded into the computational architecture. As quantum hardware evolves, architectures that reduce infeasible exploration may offer a more resource-conscious path for scaling quantum routing experiments, because they reduce the sampling effort spent on invalid configurations. These findings reinforce the central methodological argument of this paper that for transportation routing problems, the value of quantum optimization depends not only on encoding route costs, but also on how the search architecture represents and preserves routing feasibility.

## 5. Conclusions

This study develops a transportation-grounded constraint-aware QAOA$^{+}$ framework for constrained routing problems, using the TSP and VRP as canonical testbeds for sequencing, assignment, depot-return, and route-continuity requirements. The central motivation is not simply to apply a quantum algorithm to routing, but to examine how a core principle of transportation routing methodology, feasibility-preserving search, can be translated into quantum operator design. In the proposed framework, assignment-related feasibility requirements are embedded into the mixer, while travel cost, depot return, and route-continuity treatment are handled through the cost Hamiltonian.

The study compares three constraint-handling architectures: a penalty-based QUBO-QAOA formulation, a penalty-free QAOA$^{+}$ formulation with a feasibility-preserving mixer, and a Hybrid QAOA$^{+}$ formulation that combines mixer-level feasibility preservation with additional penalty guidance. The results indicate that these architectures produce different feasible-search behaviors, even on small, controlled routing instances. In particular, the QUBO-QAOA formulation tends to allocate sampling probability to infeasible or high-cost route configurations, whereas the QAOA$^{+}$ variants more effectively concentrate probability on feasible routing configurations. These findings support the argument that, for transportation routing problems, the treatment of feasibility is not only a modeling detail but a central determinant of quantum search behavior.

The comparison also suggests that the value of penalty-based guidance depends on the structure of the routing problem. For the compact single-vehicle TSP instance, the Hybrid QAOA$^{+}$ formulation improved concentration around the known optimum, suggesting that moderate penalty guidance can help intensify search within a relatively small feasible space. For the multi-vehicle VRP instance, the penalty-free QAOA$^{+}$ formulation was more robust, indicating that broader exploration within the assignment-feasible space can be more beneficial when feasible solutions differ by both route ordering and customer-to-vehicle allocation. This contrast reinforces the transportation-centered interpretation of the results implying effective quantum routing design should be sensitive to the structure of the feasible routing space, rather than relying on a single generic constraint-handling strategy.

The methodological contribution of this work is a constraint-handling architecture for a quantum routing algorithm that embeds classical feasibility-preserving logic directly into the quantum mixer, thereby enforcing feasibility through admissible search transitions rather than penalty calibration. The proposed mixer does not simply modify a generic quantum circuit, but it builds routing feasibility into the search dynamics by defining the allowable transitions between feasible route configurations and restricting evolution to valid solutions. Methodologically, this creates a superposition-based analogue of neighborhood search, where probability mass can be redistributed across multiple feasible neighbors prior to measurement while remaining within the feasible routing subspace.

The computational experiments are intentionally limited to small TSP and VRP instances in a classical quantum simulator. Accordingly, the results should not be interpreted as evidence of large-scale quantum advantage or practical scalability. Instead, they provide controlled methodological evidence on how constraint-handling architecture affects feasible-space exploration, convergence behavior, and probability concentration over high-quality routes within the quantum search. Future work can extend this framework to richer transportation routing settings, including capacity, time-window, and service constraints, and can examine how hardware noise, qubit connectivity, and circuit depth affect the feasibility-preserving properties of the proposed mixer on near-term quantum devices.

**Funding**
This material is based upon work supported by the National Science Foundation under Grant Number 2520130. Any opinions, findings, and conclusions or recommendations expressed in this material are those of the authors and do not necessarily reflect the views of the National Science Foundation.